\documentclass[]{elsart}
\usepackage{graphicx}

\usepackage{amssymb}

\usepackage{amsmath}

\usepackage{subfigure}

\begin{document}

\begin{frontmatter}

\title{Development of a triple GEM UV-photon detector
operated in pure CF$_4$ for the PHENIX experiment. }
\vspace{-12mm}
\author[Weizmann]{A. Kozlov}, 
\author[Weizmann]{I. Ravinovich}, 
\author[Weizmann,Budker]{L. Shekhtman}, 
\author[Weizmann]{Z. Fraenkel}, 
\author[Tokyo]{M. Inuzuka}, 
\author[Weizmann,corr]{I. Tserruya}

\thanks[corr]{Corresponding author: I. Tserruya, 
  Department of Particle Physics,
  Weizmann Institute of Science, Rehovot 76100 Israel.
  Tel: + 972-8-934 4052; Fax: + 972-8-934 6021.
  \it{E-mail}: itzhak.tserruya@weizmann.ac.il
}
\address[Weizmann]{ Weizmann Institute of Science, Rehovot 76100, Israel }
\address[Budker]{on leave from Budker Institute of Nuclear
  Physics, Novosibirsk 630090, Russia}
\address[Tokyo] { University of Tokyo, Tokyo 113-0033, Japan }
\maketitle

\begin{abstract}
 Results obtained with a triple GEM detector operated in pure CF$_4$ with and
 without a reflective CsI photocathode are presented. The detector operates in a
 stable mode at gains up to 10$^4$. A deviation from exponential growth starts
 to develop when the total charge exceeds $\sim 4\times10^6$ e leading to gain
 saturation when the total charge is $\sim 2\times10^7$ e and making the structure
 relatively robust against discharges. No aging effects are observed in the GEM
 foils after a total accumulated charge of $\sim 10$ mC/cm$^2$ at the anode. The ion
 back-flow current to the reflective photocathode is comparable to the electron
 current to the anode. However, no significant degradation of the CsI photocathode is observed
 for a total ion back-flow charge of $\sim7$ mC/cm$^2$.
\end{abstract}

\end{frontmatter}

\section{Introduction}
\vspace{-5mm}
A Hadron Blind Detector (HBD) is being considered for an upgrade  
of the PHENIX detector 
at the Relativistic Heavy Ion Collider (RHIC) at BNL~\cite{HBD}. 
The HBD will allow the measurement of electron-positron pairs from 
the decay of the light vector mesons, $\rho$, $\omega$ and $\phi$ 
and the low-mass pair continuum (m$_{ee}\leq$ 1 GeV/c$^2$) in 
Au-Au collisions at energies up to $\sqrt{s_{NN}} =$ 200 GeV. From 
Monte Carlo simulations and general considerations, the main HBD 
specifications are: electron identification with very high 
efficiency ($>$90\%), double hit recognition better than 90\%, 
moderate pion rejection factor of about 200, and radiation budget 
of the order of 1\% of a radiation length. The primary choice 
under study is a windowless Cherenkov detector, operated in pure 
CF$_4$ in a special proximity focus configuration, with a 
reflective CsI photocathode and a triple Gas Electron Multiplier
(GEM)~\cite{Sauli_GEM}  
detector element with a pad readout.

The proposed scheme is significantly different from other HBD
designs~\cite{giom,pisani}. The combination of a windowless detector with a CsI photocathode  
and CF$_4$ results in a very large bandwidth 
(from 6 to 11.5 eV) and a very high figure of merit N$_0$ = 940. 
With these unprecedented numbers, one expects approximately 
40 detected photo-electrons per incident electron in a 50 cm long  
radiator, thus 
ensuring the necessary high levels of single electron detection efficiency 
and double hit recognition. The scheme foresees the detection of the 
Cherenkov photoelectrons in a pad plane with the pad size approximately 
equal to the photoelectron space distribution ($\sim$10 cm$^2$). This results in a low 
granularity detector. In addition to that, since the photoelectrons 
produced by a single electron will be distributed between at most three pads,
one can expect a primary charge of at least 10 e per pad 
allowing to operate the detector at a relatively moderate gain of 
a few times 10$^3$.

In this paper, we report on the operation in pure CF$_4$ of  
a triple GEM detector, with 
and without a CsI photocathode evaporated on the top face of the 
first GEM. Extensive studies using 3$\times$3 and 10$\times$10 cm$^2$ detectors 
have been performed using a Hg UV lamp, an Fe$^{55}$ X-ray source 
and an Am$^{241}$ alpha source. All measurements were also 
performed with the conventional Ar/CO$_2$ (70/30\%) gas mixture 
for comparison. Section 2 describes the various setups and 
conditions under which the measurements were performed. The 
studies include measurements of the gain amplification curve of the 
triple GEM structure without (section 3) and with a reflective CsI 
photocathode (section 5), discharge probability in the presence 
of the high ionization induced by the Am$^{241}$ $\alpha$-particles 
(section 4) and ion back-flow to the photocathode (section 6). A 
short summary and conclusions are presented at the end of the paper 
in section 7.
\vspace{-5mm}
\section{Setup and experimental conditions. }
\vspace{-5mm}
For all the measurements, GEMs produced at CERN were used with 
50 $\mu$m kapton thickness, 5 $\mu$m thick copper layers, 60-80 
$\mu$m diameter holes and 140 $\mu$m pitch. The GEMs had 3$\times$3 or
10$\times$10 cm$^2$ sensitive areas. These two types of GEMs will be 
referred to in the text as "small" and "large" respectively. Three GEMs 
were assembled in one stack with G10 frames as shown in Fig. 1. 
The distance between the GEMs was 1.5 mm and the distance between the bottom GEM (GEM3) 
and the printed circuit board (PCB) was in most (some) cases 2 mm (1.5 mm). The
distance between the top GEM 
(GEM1) and the drift mesh was 3 mm in the measurements with X-rays and 
$\alpha$-particles and 1.5 mm in the measurements with UV-photons.

The PCB consisted of 5 strips of 100$\times$20 mm$^2$ 
each. The central group was connected either to a charge sensitive 
pre-amplifier and shaper or to a picoammeter, depending on the particular 
measurement. The other groups were grounded.

Two gases were used for the measurements in this work: 
an Ar/CO$_2$ (70/30\%) mixture and CF$_4$. We used a premixed bottle of 
Ar/CO$_2$ with Ar of 99.999\% purity and CO$_2$ of 99.998\% 
purity. The purity of the CF$_4$ was 99.999\%.

\begin{figure}[htbp]
 
  \centering
  \includegraphics[keepaspectratio=true, width=12cm]{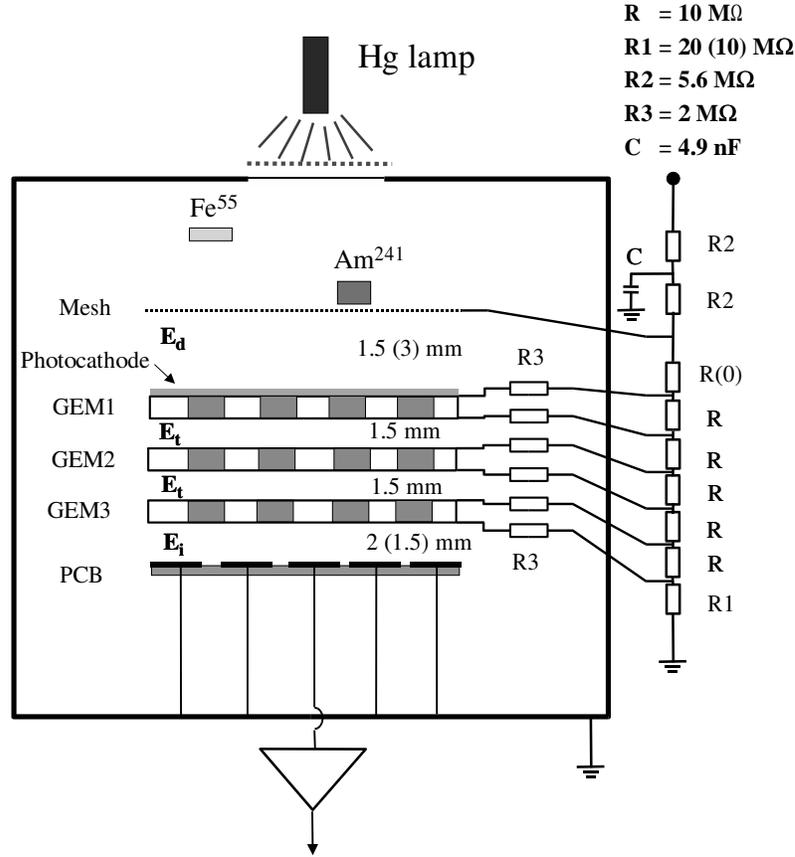}
  
  \caption{Setup of the triple GEM detector and resistor chain. The Hg lamp,
    Fe$^{55}$ and Am$^{241}$ sources were used for measurements with UV-photons,
    X-rays and $\alpha$-particles, respectively.}
  \label{fig:rch}

\end{figure}

High voltage was supplied to the GEM  electrodes via a resistive chain (see Fig.~1). 
For most of the measurements, the resistors R were equal to 10 M$\Omega$ whereas
the resistor R1 
feeding the gap between GEM3 and PCB (see Fig.1)  was equal to 20 M$\Omega$. In
some measurements with X-rays and $\alpha$-particles R1 was equal to 10
M$\Omega$. For some measurements an independent voltage supply for selected electrodes was used. 
In particular, this was required for the measurement of the ion current to 
the top electrode of GEM1, while studying the ion back-flow (see 
below).

We use the gap names and the field notations as proposed in~\cite{Sauli_tr},
i.e. the gap between the mesh  
and top GEM is called "drift" and the corresponding field is referred 
to as E$_d$; the gaps between GEMs are called "transfer" and 
the corresponding fields are referred to as E$_t$; the gap between 
GEM3 and the PCB is called "induction" and the corresponding field is 
referred to as E$_i$. Most measurements were performed with a 2 mm induction gap
and a 20 M$\Omega$ resistor feeding it. In this configuration, when the voltage
across the GEMs is 510 (370) V, corresponding to a gain of $\sim10^4$ in CF$_4$
(Ar/CO$_2$), the transfer and induction fields are about 3.4 (2.5) kV/cm and 5.1
(3.7) kV/cm, respectively. When R1 is equal to 10 M$\Omega$, the induction fields
are half the quoted values. The ability of the GEM to 
transport electrons through its holes is referred to as "electron
transparency". It is the product of two factors: the fraction 
of electrons collected from the top gap into the holes and the 
fraction of electrons extracted from the holes into the bottom gap. 
The electron transparency of the GEMs with the voltages and fields  
indicated above, can be derived from the data presented in 
\cite{Sauli_tr}. For GEM1 and GEM2 the electron transparency is close 
to 1, while for GEM3 it is about 0.7 in the case of the lower induction 
field and approaches 1 for the high induction field.

The photocathode was prepared by  evaporating a $\sim 2000$~\AA~ thick layer of 
CsI on the first GEM previously coated with thin layers of Ni and Au to avoid
chemical interaction with the CsI film.
For the operation with the reflective photocathode the drift field has to be 
zero or even reversed in order to collect all the photo-electrons 
from the CsI layer~\cite{dirk}. For those measurements the corresponding resistor 
in the chain was shorted. The measurements with the CsI 
reflective photocathode were performed with a Hg lamp and  
a UV-transparent window (CaF$_2$) in the cover of the detector box. The lamp  
was positioned  at the detector window with an absorber that  
reduced the UV flux $\sim$ 1000 times to avoid possible damage of the  
photocathode~\cite{singh}. The illuminated area of the detector was about 100 
mm$^2$. In this geometry, the measured photo-electron current  was 
about 2$\times$10$^6$ e/(mm$^2\times$s).

The detector assembly (drift mesh, triple-GEM, and PCB) were mounted in a stainless steel box 
that could be pumped down to $10^{-6}$ torr and was connected to 
the inlet and outlet gas lines to allow gas flushing. All measurements were done
at atmospheric pressure with an overpressure of 0.5 torr in the detector vessel. The system 
contained also devices for the precise measurement of 
temperature, pressure and water content down to the  ppm level.  
The Fe$^{55}$ X-ray source was 
positioned inside the box at a distance of $\sim$ 40 mm from the 
mesh. The total rate of X-rays was kept at the level of 1 kHz. 5.9 keV 
photons from Fe$^{55}$ release 210 e in Ar/CO$_2$ (26 eV 
per electron-ion pair) and 110 e in CF$_4$ (54 eV per  
electron-ion pair) \cite{El_numbers}.

The discharge limit in the presence of heavily ionizing particles was studied with an 
Am$^{241}$ source that emits 5.5 MeV $\alpha$-particles. The source 
in a container was attached directly to the drift mesh and 
strongly collimated in order to provide high energy deposition and 
small energy dispersion in the drift gap. The rate of the 
$\alpha$-particles varied between $100 - 300$ Hz. The distance between 
the active surface of the source and the drift mesh was $\sim$10 mm.  
The range of 5.5 MeV $\alpha$-particles in Ar/CO$_2$ is $\sim$ 39 mm  
and about 18 mm in CF$_4$. Assuming perpendicular incidence 
of the $\alpha$-particles to the drift gap, the energy deposition in a 3 mm gas
layer is estimated to be $\sim$ 1.1 MeV for 
CF$_4$ and $\sim$ 0.30 MeV for Ar/CO$_2$ producing $\sim$ 20000 
and $\sim$ 12000 primary charges, respectively.

For the study of gain limits we needed a reliable way to monitor 
the discharges in the triple GEM assembly. The resistor 
chain voltage was supplied by a HV power supply CAEN N126. This module allowed  
us to install a protection against over-current with a precision of 
0.1 $\mu$A. The protection threshold was always kept at 
5 $\mu$A above the chain current which was usually in the range 
between 50 and 100 $\mu$A. This was enough to cause a trip 
when a discharge occurred in a single GEM. The trip signal was 
reset after 1 second and counted by a scaler. 
\vspace{-5mm}
\section {Gain in Ar/CO$_2$ and CF$_4$.}
\vspace{-5mm}
The gain as a function of the voltage across the GEM ($\Delta$V$_{GEM}$) 
was measured with all GEMs  at the same voltages for both Ar/CO$_2$ 
and pure CF$_4$. The absolute gas gain was determined from the 
measurements of the signal from Fe$^{55}$ 5.9 keV X-ray photons. An 
example of the pulse height spectrum for both gases is shown 
in Fig. 2. For Ar/CO$_2$ the main peak is very well 
separated from the escape peak of Ar and the energy resolution is 
$\sim22$\% FWHM. For CF$_4$ the energy resolution is close to 38\% FWHM.
In both cases the pulse height spectra were measured at a gain higher 
than $10^4$.

\begin{figure}[htbp]
  \vspace{-2mm}
  \centering
  \includegraphics[keepaspectratio=true, width = 9cm]{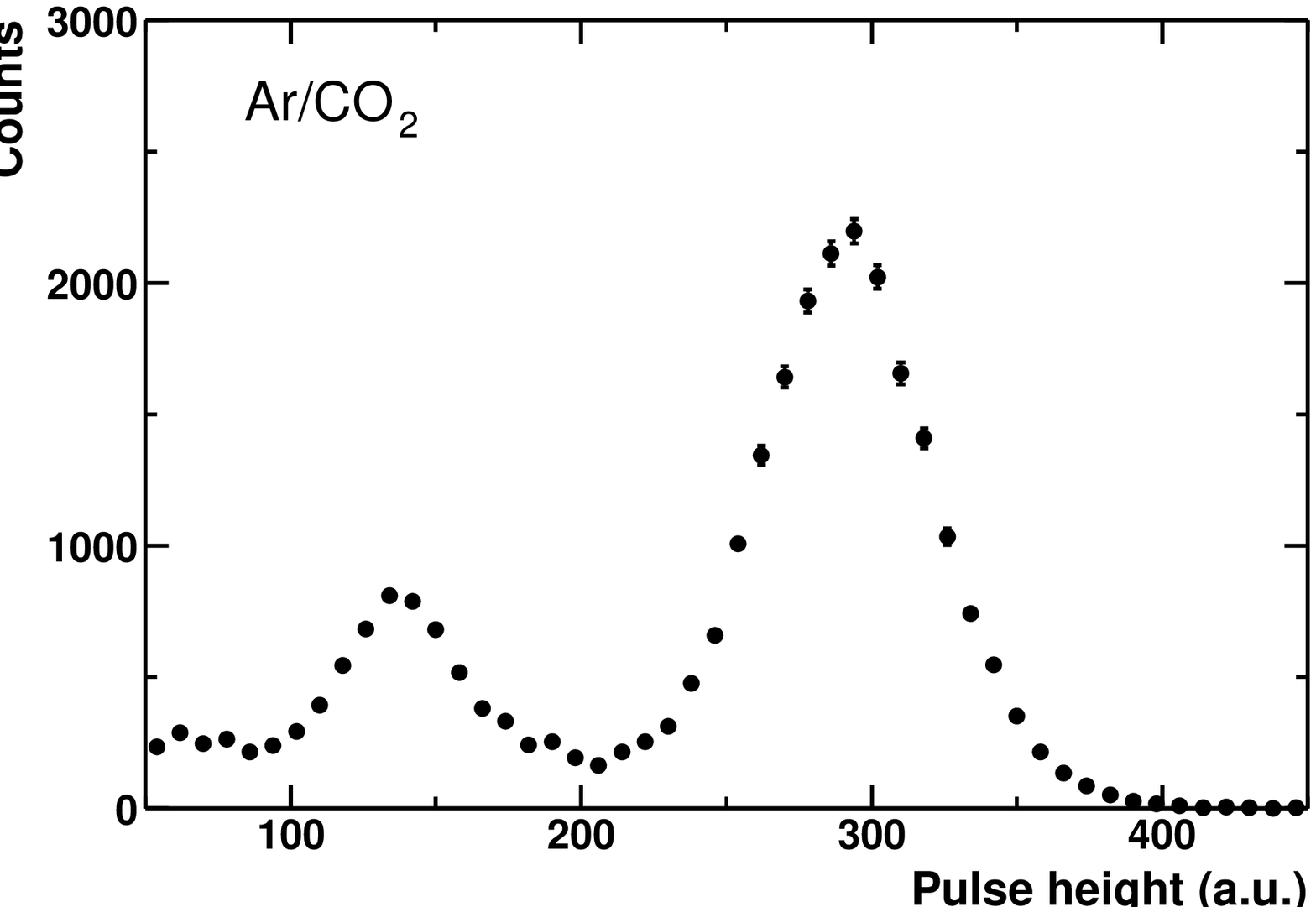}
  \hspace{-2cm}
  \vspace{-3mm}
  \includegraphics[keepaspectratio=true, width = 9cm]{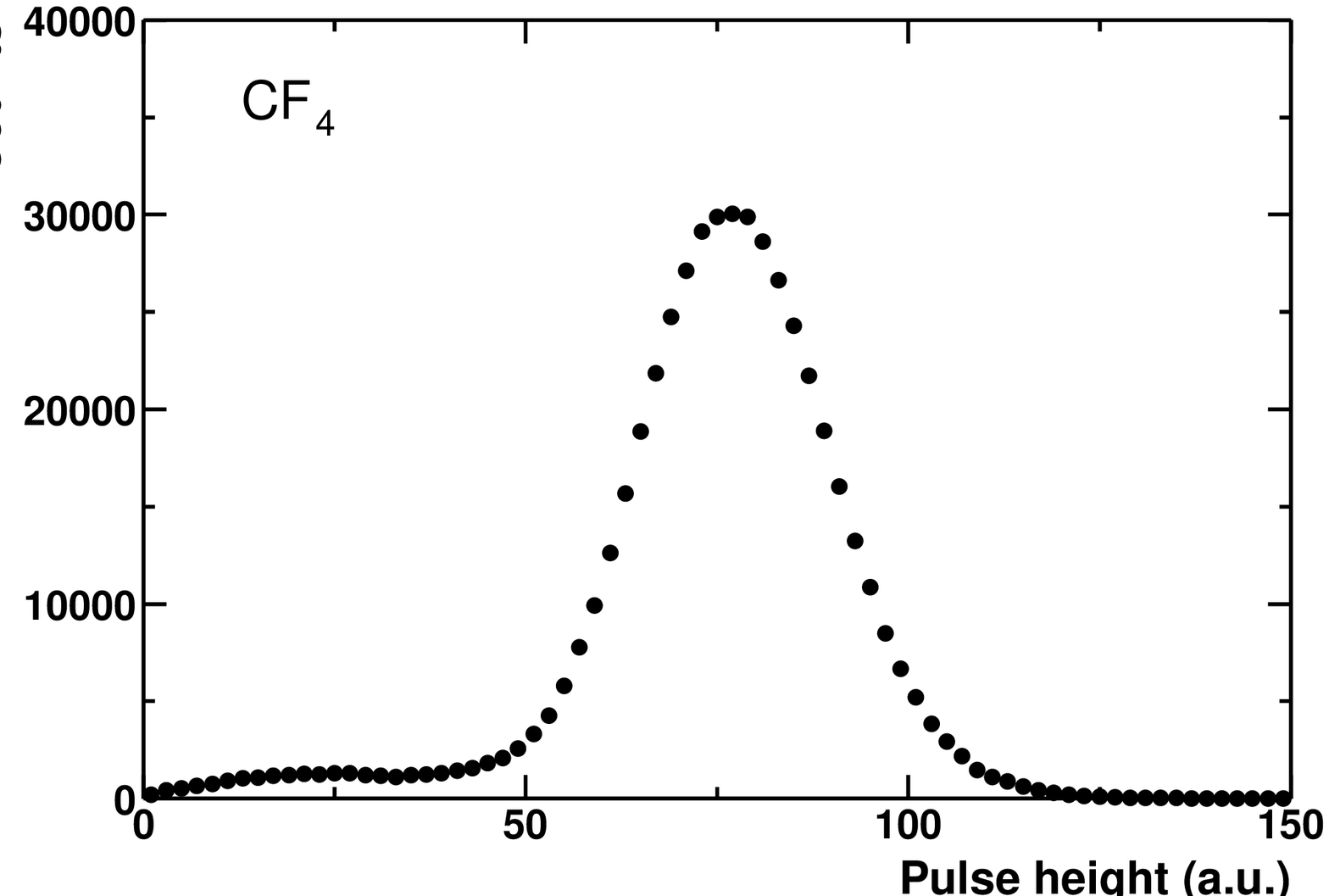}
  \caption{Pulse height spectrum of Fe$^{55}$ X-rays: top panel with Ar/CO$_2$
    (70/30\%) and bottom panel with CF$_4$. } 
\end{figure}

The gain was calculated, using the measured relationship between the output 
signal from the amplifier and the input charge to a calibration capacitor 
and taking into account the average charge produced by one  
5.9 keV photon (see previous section). 

Fig. 3 shows the typical gain curves measured with 5.9 keV X-rays in Ar/CO$_2$ and CF$_4$ using small and
large GEMs. Several detector sets were used and good reproducibility between the
various sets was observed. Comparing the data 
for Ar/CO$_2$ and CF$_4$ in Fig. 3 one can see that the operational 
voltage for CF$_4$ is $\sim$140 V higher but the slopes of the  
gain-voltage 
characteristics are similar for both gases, i.e. an increase of 20 V  
in $\Delta$V$_{GEM}$ causes an increase of the gain by a factor of  
$\sim3$. The gain in CF$_4$ can reach values above $10^5$, in 
spite of the very high operational voltage, as was already reported in 
\cite{Amos1}.

\begin{figure}[htbp]
  \centering
  \includegraphics[keepaspectratio=true, width = 10cm]{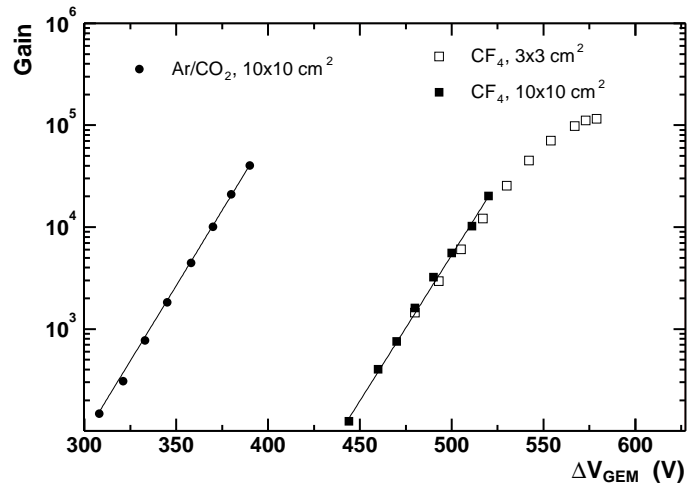}
  \caption{Gain as a function of GEM voltage measured with Fe$^{55}$ X-ray
    source. The 3$\times$3 cm$^2$ detector had a 
    CsI layer deposited on
    the top face of GEM1. The lines represent exponential fits to the data with
    10$\times$10 cm$^2$ GEMs.}
\end{figure}

The absolute value of the gain is very sensitive to the gas density. Small
variations of the gas pressure ($P$) and/or temperature ($T$) significantly affect
the gain as demonstrated in Fig. 4. A change of 1$\%$ in the $P/T$ value causes
a gain variation of 17$\%$ in Ar/CO$_2$ and of 26$\%$ in CF$_4$.

\begin{figure}[htbp]
  \centering
  \includegraphics[keepaspectratio=true, width = 10cm]{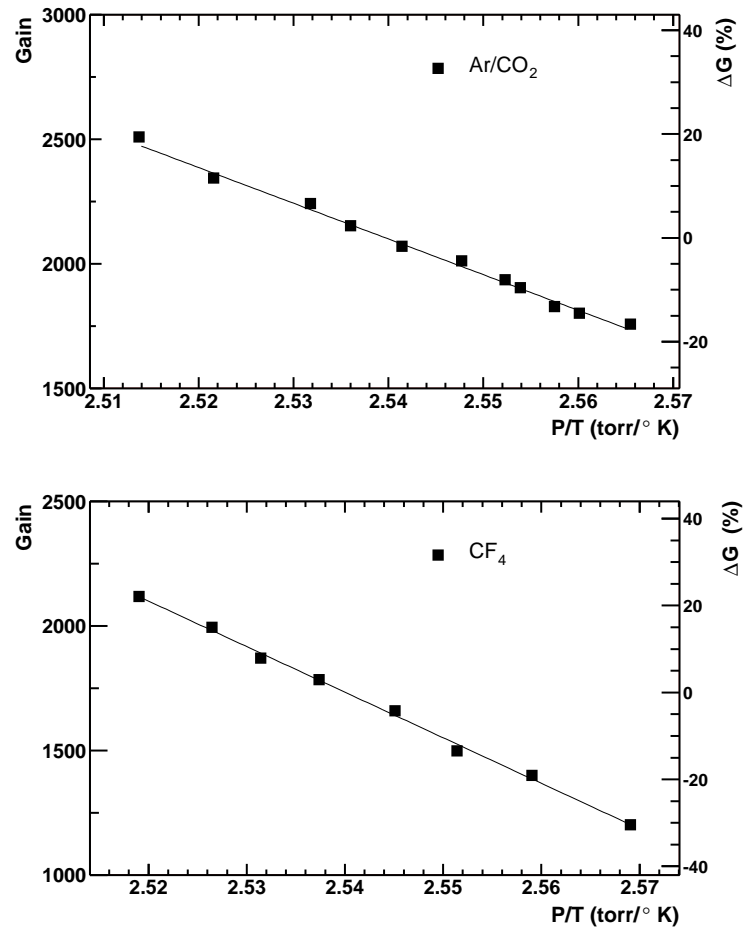}
  \caption{Effect of gas density on the gain in Ar/CO$_2$ (top panel) and pure
    CF$_4$ (bottom panel). The relative gain variation ($\Delta$G) is calculated
    with respect to the gain at $P/T$ = 2.54 torr/$^{\circ}$K.}
\end{figure}

Another feature of CF$_4$ which can be seen in Fig. 3 is the strong 
deviation from exponential growth at high gains. This ``non-linearity'' is much more 
pronounced when the detector is irradiated with Am$^{241}$ 
$\alpha$-particles (Fig. 5). In that figure the saturation level of the 
pre-amplifier is marked with a dashed line. In the case of Ar/CO$_2$ 
the charge depends on $\Delta$V$_{GEM}$ exponentially, and the 
signal is saturated by the pre-amplifier. In pure CF$_4$, on the other 
hand, the 
dependence of charge versus $\Delta$V$_{GEM}$ becomes non-linear 
above the value of $\sim4\times10^6$ e and is completely saturated at 
$\sim2\times10^7$ e, which is below the saturation level of the 
pre-amplifier. This difference in performance in Ar/CO$_2$ and 
pure CF$_4$ may be due to the higher primary charge density 
and lower diffusion in CF$_4$. These two features make the charge 
cluster in CF$_4$ more compact and dense and, as a consequence, 
increase the electric field inside the charge cloud resulting  
in the saturation of the avalanche. This saturation effect is of prime
importance for the anticipated application of the HBD in the PHENIX experiment
where single photoelectrons are to be detected in a high multiplicity environment
of charged particles.

\begin{figure}[htbp]
  \centering
  \includegraphics[keepaspectratio=true, width = 10cm]{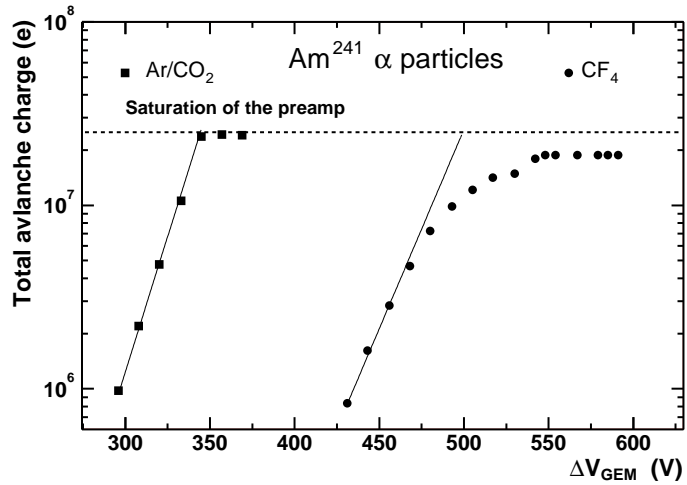}
  \caption{Total avalanche charge as a function of GEM voltage measured with
    Am$^{241}$ $\alpha$-particles. The lines 
    represent exponential growth of the total charge in the avalanche derived from
    the low gain points.}
\end{figure}

\vspace{-5mm}
\section {Discharge probability in the presence of heavily ionizing particles.}
\vspace{-5mm}
Stability of operation and absence of discharges in the presence 
of heavily ionizing particles is crucial for the operation of the 
HBD. The Am$^{241}$ source was used to simulate heavily ionizing 
particles under laboratory conditions. We 
determined quantitatively the probability of discharge as the ratio 
between the number of discharges within a certain period of time 
and the number of $\alpha$-particles traversing the detector during 
the same period. The discharge 
probability was measured in small GEMs and the results are shown in 
Figs. 6a and 6b in two different forms: 
as a function of GEM voltage and as a function of gain. 

For the Ar/CO$_2$ mixture the probability of discharge exhibits a rapid 
increase between 400 V and 420 V across the GEM when the gain 
reaches $3\times10^4$. In terms of gain and GEM voltage these results agree 
with similar data from \cite{Sauli1}. In CF$_4$ the discharge 
probability grows at $\Delta$V$_{GEM}$ above 590 V  
with both E$_i$ = 2.6 kV/cm and E$_i$ = 5.1 kV/cm. The second setup 
also had a CsI photocathode on GEM1. From Fig. 5 one can see that the 
signal from $\alpha$-particles in CF$_4$ is completely saturated  
above $\Delta$V$_{GEM} \sim540$ V at the level of $\sim2\times10^7$ e. As a 
consequence, the total charge produced by the heavily ionizing particle is 
limited to below the Raether limit~\cite{raeth} and its ability to provoke a discharge 
is strongly suppressed. Thus, the gain in 
CF$_4$ even in the presence of $\alpha$-particles can reach extremely 
high values of close to $10^6$. The HBD is expected to operate at gains 
$\le 10^4$, i.e. with a comfortable margin below  
the discharge threshold. 

\begin{figure}[htbp]
  \centering 
  \includegraphics[keepaspectratio=true,width = 10cm]{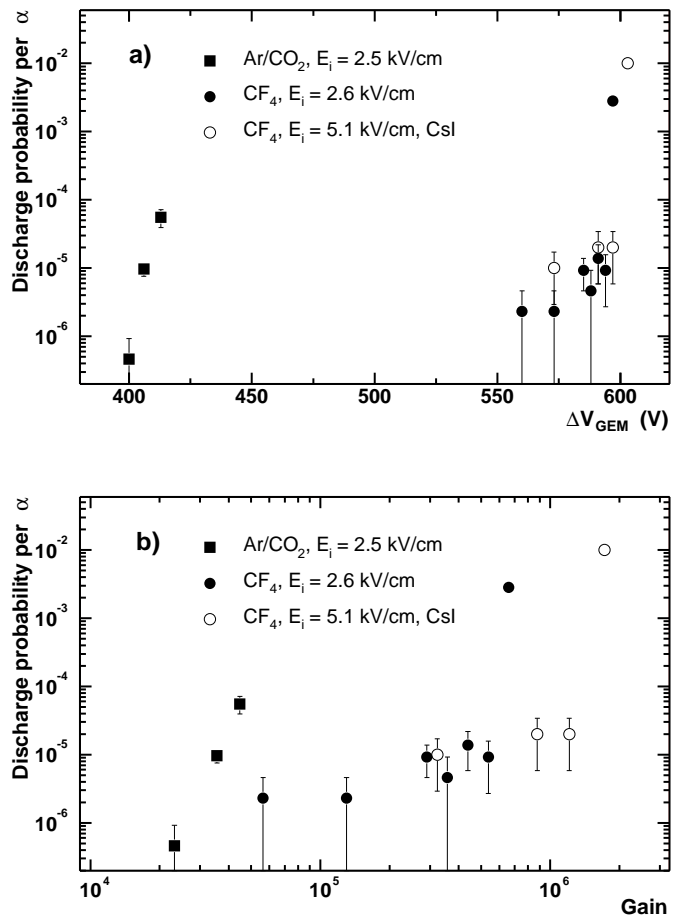}
  \caption{Discharge probability per $\alpha$-particle as a function of: a) GEM
    voltage; b) gain. The values of the induction field E$_i$ refer to a gain of
    10$^4$. The error bars represent the statistical error. The two highest points
    for CF$_4$ represent a lower limit of the discharge probability.} 
\end{figure}

The measurements of the discharge probability were also performed with 
the large GEM setup. However during the measurement in CF$_4$ the 
GEMs were severely damaged by the very first spark and a similar 
study could not be conducted for this setup. The damage to the GEM  
was severe due to 
the combination of high operational voltage and high capacitance 
which results in the energy deposited in the discharge being too 
high. We plan to repeat the studies with large GEMs with a 
proper segmentation of the GEMs so as to reduce their capacitance.  

\section{ Operation with the CsI reflective photocathode.}

In all the tests with the CsI photocathode a mercury lamp was used for 
irradiation. In order to determine the total emission from the 
photocathode itself without any 
amplification in the GEMs, we applied a positive voltage between GEM1 and the 
mesh, thus collecting the emitted photo-electrons in the mesh. The 
operation of the CsI photocathode is shown in Fig. 7, where the 
photo-electron current as a function of voltage (7a) and time 
(7b) is plotted. From Fig. 7a it is seen that in order to measure  
the full photo-electron emission the voltage between the 
mesh and GEM1 has to exceed  200 V or, since the drift gap was 1.5 mm, the field has
to be higher than 1.3 kV/cm, in agreement with~\cite{Amos1}.

\begin{figure}[htbp]
  \centering 
  \includegraphics[keepaspectratio=true, width = 10cm]{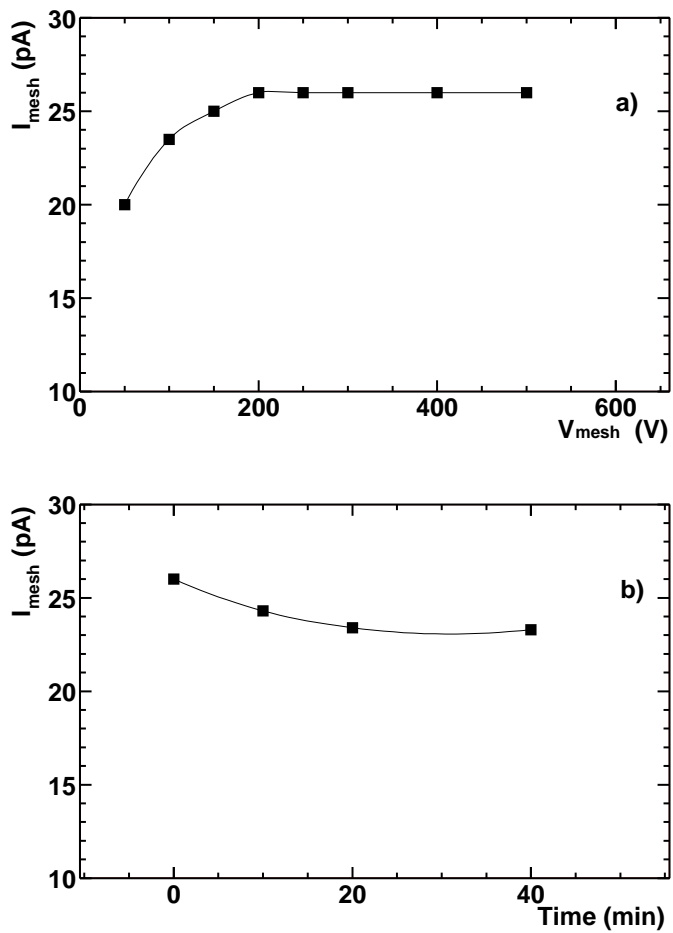}
  \caption{Current from GEM1 to the mesh: a) as a function of voltage; b) as a
    function of time. The lines are to guide the eye.} 
\end{figure}

In Fig. 7b the value of the current to the mesh as a function of time 
is shown, demonstrating that one has to wait about 30 min after the 
application of the HV in 
order to stabilize the signal. As CsI is a semi-insulating material, 
this initial instability of the signal might be caused by polarization 
and up-charging of the layer. 

The study of the triple GEM detector with a reflective 
photocathode was always performed in the regime with E$_d$ = 0. 
Fig. 8 shows the current to the PCB as a function of 
the GEM voltage for the small GEM setup. The measurements were 
done in Ar/CO$_2$ and CF$_4$. In the CF$_4$ curve we 
can clearly see two regions well described by two exponential dependencies on
$\Delta$V$_{GEM}$ (see lines in Fig. 8): an initial slow increase of current at lower voltages related
to the increase of the extraction of the photo-electrons from the CsI surface
into the holes of GEM1 and a steep 
exponential increase at higher voltages due to amplification in the 
GEMs. A detailed discussion of these processes and the transition from one region to 
the other can be found in \cite{Amos2}. In Ar/CO$_2$ these two regions  
are not so clearly separated because amplification in this mixture 
starts at lower voltages. The electron extraction cannot exceed the maximum level shown in 
Fig. 7a. It indeed seems to reach this level of 100\% extraction  
indicated by the dashed line in Fig. 8. Thus, the gain is determined as the ratio between the 
current to the PCB and the the extraction current. The latter is given by the
first exponential curve up to $\Delta$V$_{GEM}$ = 350 V and by the 100\% extraction
value at higher values of $\Delta$V$_{GEM}$.

\begin{figure}[htbp]
  \centering
  \includegraphics[keepaspectratio=true, width = 10cm]{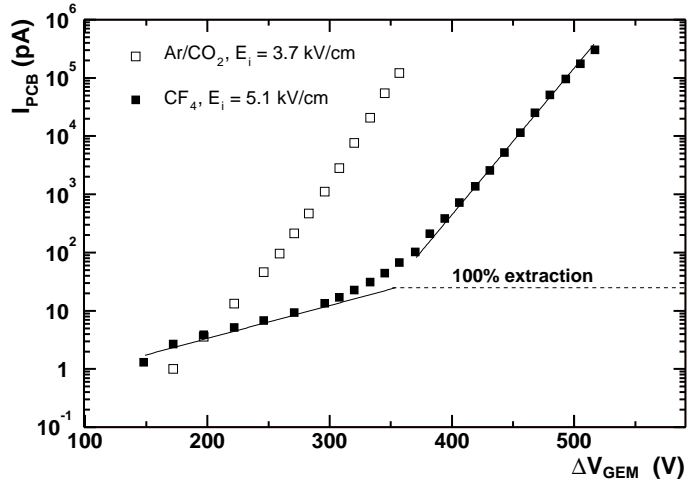}
  \caption{Current to the PCB as a function of $\Delta$V$_{GEM}$.}
\end{figure}

The gain as a function of $\Delta$V$_{GEM}$ for the setup with the 
reflective photocathode is shown in Fig. 9. In the same figure the 
data obtained with X-ray irradiation (Fe$^{55}$) are also shown in order to 
demonstrate that the different methods of gain measurement give similar 
results.

\begin{figure}[htbp]
  \centering
  \includegraphics[keepaspectratio=true, width = 10cm]{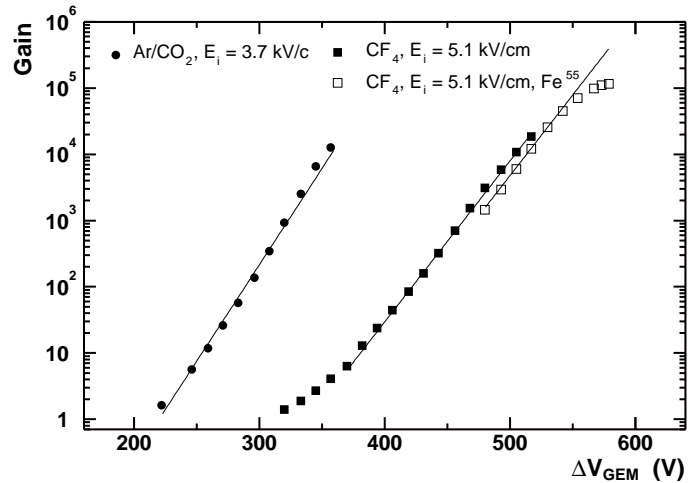}
  \caption{Gain as a function of $\Delta$V$_{GEM}$ for Ar/CO$_2$ and CF$_4$
    measured with the UV lamp. For CF$_4$, the gain curve with Fe$^{55}$
    is also shown. The lines are exponential fits to the data.}
\end{figure}

\section{Ion back-flow in the triple GEM detector operating with a reflective
  photocathode.} 

The flow of positive ions to the CsI layer is one of the
potential damaging factors that can cause aging of the photocathode
\cite{singh,Amos3,Nov1,Sauli2}. We call this factor ion back-flow and
characterize it by the ratio between the current to the top electrode
of GEM1 and the current to the PCB. This ratio depends on both the ion
current itself and the fraction of electron current flowing to the
PCB. This is a convenient definition as
it allows us to estimate the actual ion current from the measured signal
at the PCB. In order to
measure the current to the photocathode we supplied the voltage separately
to the top electrode of GEM1 with a CAEN N126 power supply.
The voltages to all other electrodes were supplied through the resistive chain.

In Fig. 10 the ratio of the current to the photocathode and the current
to the PCB (ion back-flow factor) as a function of gain is shown
for different conditions. The errors on the plots
are mainly due to the limited accuracy of the photocathode current
measurements. The value of the induction field was changed by changing the
corresponding resistor in the chain and the value, indicated 
in the caption (5.1 kV/cm), is reached at a gain of $10^4$. 

\begin{figure}[htbp]
  \centering
  \includegraphics[keepaspectratio=true, width = 11cm]{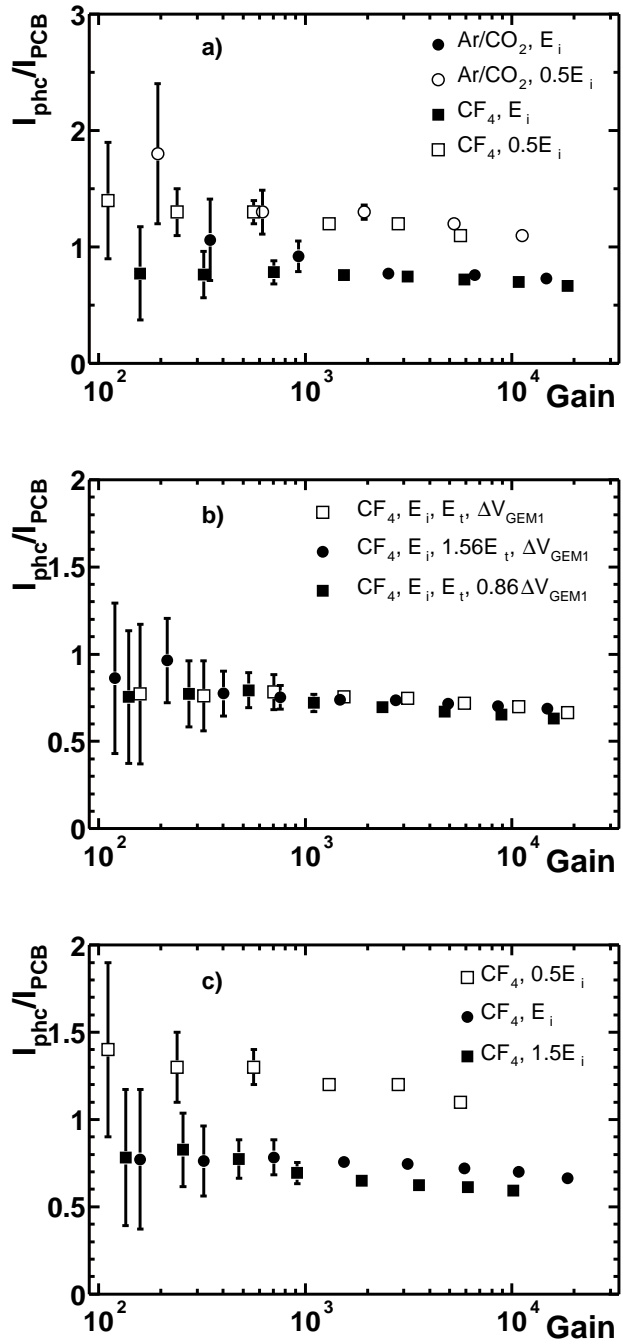}
  \caption{Ion back-flow factor as a function of gain. a) Comparison of ion
    back-flow factor for Ar/CO$_2$ and CF$_4$ and two different induction fields:
    standard E$_i$ = 5.1~kV/cm and 0.5~E$_i$. The values refer to a gain of
    $10^4$; b) Ion back-flow factor for different 
    electrostatic conditions in the region between GEM1 and GEM3. c) Ion back-flow
    factor for 3 different values of the induction field.} 
\end{figure}

In Fig. 10a we see that in spite of the very different transport
properties of the gases used in the measurements no significant dependence of
the ion back-flow factor on the nature of the gas is observed as a function of gain and
for different induction fields. The insensitivity of the ion
back-flow factor 
to the particular gas at moderate gains is similar to that seen in
\cite{Nov1}. It means that the efficiency of the transport of
electrons and ions through the GEMs is the same for both
gases and does not depend on diffusion.

The insensitivity of the ion back-flow factor to the electric field
between the GEMs and in the GEM is demonstrated in Fig. 10b. Here the value of the
ion back-flow factor as a function of gain 
is shown for three different electrostatic conditions: 1) standard,
when the transfer field is equal to 3.4 kV/cm for both gaps and
the induction field is equal to 5.1 kV/cm (the values refer to a 
gain of 10$^4$), 2) enhanced transfer field in both gaps, 3) reduced field in
GEM1. 
From Fig. 10b we see that neither variation in electrostatic
conditions between nor inside the GEMs affect significantly the ion
back-flow factor. 

The only parameter which affects the value of the ion back-flow
in our case is the induction field. Fig. 10c shows the value of the ion
back-flow factor as a function of the gain for 3 values of the induction
field. The field in the induction gap does not affect the
ion flow itself as ions are produced in the holes of the last GEM or
in their vicinity, collected into the holes and then
transported to the top gap. The only factor that is affected is
the electron flow from  GEM3 to the PCB. Thus the 
ion back-flow factor being higher than one at low induction field
means that a fraction of the electrons is collected at the bottom face of GEM3
and consequently the amount of ions reaching the photocathode can be larger than the amount of
electrons collected at the PCB. The increase of the
induction field improves the electron collection efficiency at the PCB and
reduces the value of the ion back-flow factor. It is clear from the
figure that for $E_{i}$ above 5 kV/cm the collection efficiency does
not increase significantly resulting in a minimum value of the ion
back-flow factor of $\sim0.7$ at a gain of $10^4$, consistent with results of~\cite{Amos3}.

During these measurements the photocathode was exposed to a total
ion charge of $\sim$ 7 mC/cm$^2$. This charge density
corresponds to $\sim10$ hours of continuous irradiation with 
$\sim10^7$ photons/(mm$^2\times$s) at a gain of $10^4$. In spite of this quite high ion
back-flow the CsI quantum efficiency loss was not more than 30\% after this 
irradiation.

\section{Summary and conclusions}

We have presented very encouraging results on the operation of 
a triple GEM detector
in pure CF$_4$ with and without a reflective CsI photocathode. The
slope of the gain curve is similar to that of the conventional
Ar/CO$_2$ (70/30\%) gas mixture, however $\sim$ 140 V higher voltage
across the GEMs is needed for a given gain. The gain curve starts
deviating from exponential growth when the total charge in the
detector exceeds $\sim 4 \times 10^6$ e, and the gain is fully saturated when the
total avalanche charge reaches $\sim 2 \times 10^7$ e. This is an interesting
property making the system more 
robust against discharges as compared to Ar/CO$_2$. Stable
operation can be achieved at gains up to 10$^4$ in the presence of
heavily ionizing particles. No deterioration  of the GEM foil performance in a
pure CF$_4$ atmosphere was observed for a total accumulated 
charge of $\sim$ 10 mC/cm$^2$ at the PCB. The ion back-flow to the photocathode is close to
100\%, independent of the operating gas and of the transfer field E$_t$ between
successive GEMs. At a gain of 10$^4$, the ion back-flow factor can be reduced to
$\sim 70\%$ by applying a relatively high induction field of E$_i \sim $ 5
kV/cm. In spite of the high ion back-flow no sizable deterioration of the CsI quantum efficiency was
observed when the photocathode was exposed to a total ion charge of
$\sim$ 7 mC/cm$^2$. This value is larger by about two orders of magnitude than the
total integrated ion charge density expected during the lifetime of the planned
HBD.

\section{Acknowledgments}

We thank F. Sauli, A. Breskin, R. Chechik, M. Klin and D. M\"ormann for their invaluable
help and very useful discussions. This work was partially supported by the Israel Science
Foundation, the Nella and Leon Benoziyo Center of High Energy Physics Research
and the US Department of Energy under Contract No. DE-AC02-98CH10886.

\end{document}